\documentstyle[prl,twocolumn,aps]{revtex}
 \sloppypar
\begin{document}
\draft
\twocolumn
[\hsize\textwidth\columnwidth\hsize\csname
@twocolumnfalse\endcsname
\title{Multi-particle States in Spin-1 Chain System ${\bf CsNiCl_{3}}$}
\author{M. Kenzelmann$^{1}$, R.~A. Cowley$^{1}$, W.~J.~L. Buyers$^{2,3}$,
R. Coldea$^{4,5}$, J.~S. Gardner$^{2}$, M. Enderle$^{6}$, D.~F.
McMorrow$^{7}$, S.~M. Bennington$^{5}$}
\address{(1) Oxford Physics, Clarendon Laboratory, Oxford OX1 3PU, UK (2)
Neutron Program for Materials Research, National Research Council
of Canada, Chalk River, Ontario, Canada KOJ 1J0 (3) Canadian
Institute for Advanced Research (4) Oak Ridge National Laboratory,
Solid State Division, Oak Ridge, TN 37831, USA (5) ISIS Facility,
Rutherford Appleton Laboratory, Oxon OX11 0QX, UK (6) Technische
Physik, Geb. 38,  Universit\"{a}t des Saarlandes, 66123
Saarbr\"{u}cken, Germany (7) Condensed Matter Physics and
Chemistry Department, $Ris\o$ National Laboratory, DK-4000,
Roskilde, Denmark}

\date{\today}
\maketitle
\begin{abstract}
A continuum of magnetic states has been observed by neutron
scattering  from the spin-1 chain compound ${\rm CsNiCl_{3}}$ in
its disordered gapped one-dimensional phase. Results using both
triple-axis and time-of-flight spectrometers show that around the
antiferromagnetic point $Q_{\rm c}=\pi$, the continuum lies higher
in energy than the Haldane gapped excitations. At $6\;\mathrm{K}$
the integrated intensity of the continuum is about 12(2)\% of the
total spectral weight. This result is considerably larger than the
1-3\% weight predicted by the non-linear sigma model for the
3-particle continuum.
\end{abstract}


]

\newpage

The excitations of one-dimensional (1D) Heisenberg
antiferromagnets have attracted much experimental and theoretical
attention ever since Haldane \cite{F_Haldane} predicted that the
excitations of integer spin and half-integer spin chains are
different. Half-integer spin chains have no spin gap and exhibit a
spinon continuum extending to zero energy
\cite{D_Tennant,G_Muller}. For integer-spin chains the spectrum is
at low temperatures dominated by well-defined single particle
excitations corresponding to a triplet of spin-1 particles that
exhibit a large energy gap \cite{William_Buyers_1,J_Renard}. These
have been thoroughly studied by neutron scattering from
\mbox{spin-1} chain compounds and it has been shown that the
well-defined excitations largely exhaust the total scattering
\cite{Shaolong_Ma}. Nonetheless, a continuum of multi-particle
scattering was predicted by Haldane and others \cite{Haldane_2},
but as yet no experiment has established its existence. To search
for such a continuum we have measured the inelastic neutron
scattering of ${\rm CsNiCl_{3}}$ in its paramagnetic 1D phase
using time-of-flight and triple-axis spectrometers. We find a
neutron scattering continuum at energies higher than the
well-defined single particle excitation, whose integrated
intensity is bigger than predicted for the multi-particle
scattering.\par

${\rm CsNiCl_{3}}$ is magnetically a quasi 1D spin-1 chain
compound with the Ni chains along the c-axis of the hexagonal unit
cell. Each unit cell contains two Ni atoms and the Hamiltonian is
\begin{equation}
H=J \sum_{i}^{\rm chain} \vec{S}_{\rm i} \cdot \vec{S}_{\rm i+1}+
J' \sum_{<i,j>}^{\rm plane} \vec{S}_{\rm i} \cdot \vec{S}_{\rm j}
- D \sum_{i} (\vec{S_{\rm i}^{z}})^2\, . \label{Hamiltonian}
\end{equation}
The exchange interaction $J=2.28\;\mathrm{meV}$ along the $c$-axis
is much stronger than the exchange interaction in the basal plane
$J'=0.044\;\mathrm{meV}$
\cite{William_Buyers_1,Rose_Morra,Katori}. The weak Ising
anisotropy $D=4\;\mathrm{\mu eV}$ is small enough that ${\rm
CsNiCl_{3}}$ is a good example of an isotropic Heisenberg
antiferromagnet. Below $T_{\rm N}=4.85\;\mathrm{K}$ the interchain
coupling causes long-range ordering of the magnetic moments. Above
$T_{\rm N}$ ${\rm CsNiCl_{3}}$ is in a 1D magnetic phase for which
the magnetic exchange interaction along the $c$-axis dominates the
spin dynamics, except close to $T_{\rm N}$.\par

The sample of ${\rm CsNiCl_{3}}$ was a single crystal
$20\;\mathrm{mm} \times 5\;\mathrm{mm} \times 5\;\mathrm{mm}$ and
was mounted with its ($hhl$) crystallographic plane in the
principal scattering plane. Inelastic neutron scattering
experiments were carried out with the time-of-flight spectrometer,
MARI, at the ISIS facility of the Rutherford Appleton Laboratory
and with reactor-based triple-axis spectrometers, DUALSPEC at the
Chalk River Laboratories and RITA at ${\rm Ris\o}$ National
Laboratory.\par

For the experiment using MARI the incoming neutron energy was $20$
and $30\;\mathrm{meV}$. The energy resolution was $0.35$ and
$0.4\;\mathrm{meV}$, as determined from the full width at half
maximum (FWHM) of the quasi-elastic peak. The resolution in
wave-vector transfer at zero energy transfer was typically
$0.02\;\mathrm{\AA^{-1}}$ along the $c^{\star}$-axis and along the
$[110]$ direction and up to $0.19\;\mathrm{\AA^{-1}}$
perpendicular to the scattering plane if only the central detector
bank was used. Both the energy and the wave-vector resolution
improved with increasing energy transfer. The measurements were
performed with the sample at temperatures between 6.2 and
$12\;\mathrm{K}$.\par

The measurements at the DUALSPEC triple-axis spectrometer were
performed at $8.5\;\mathrm{K}$ with a fixed scattered neutron
energy of $14.51\;\mathrm{meV}$ and with a graphite filter to
absorb the higher order reflections from the pyrolytic graphite
monochromator and analyzer. The collimation from reactor to
detector was $0.65^{\rm o}$-$0.6^{\rm o}$-$1.4^{\rm o}$-$2.0^{\rm
o}$ and gave an energy resolution of $1.94\;\mathrm{meV}$, as
determined from the quasi-elastic peak. In the case of the RITA
triple-axis spectrometer, supermirror guides and a rotating
velocity selector produced a variable energy monochromatic neutron
beam with suppressed higher order contamination. The scattered
neutrons were analyzed using the Soller geometry with a fixed
energy of $5\;\mathrm{meV}$ and a cooled beryllium filter between
the sample and analyzer. The effective collimation was $1.45^{\rm
o}$-$1.45^{\rm o}$-$1^{\rm o}$-$2.0^{\rm o}$ and the energy
resolution was $0.35\;\mathrm{meV}$ as determined from the
quasi-elastic peak.\par

The measurements with MARI at temperatures between $6.2$ and
$12\;\mathrm{K}$ were performed with the c-axis of the sample
either perpendicular or parallel to the incoming beam. The dynamic
structure factor $S(\vec{Q},E)$ was measured for wave-vector
transfers along the c-axis between $l \simeq 0.175$ and $\sim
1.5$, where the wave-vector along the chain is $\frac{2\pi l}{c}$.
Because there are two Ni atoms in each unit cell along c, $l=1$
corresponds to the antiferromagnetic (AF) wave-vector, $\pi$,
generally used in theoretical work on AF chains. The data with $l
> 1.5$ was contaminated by phonon scattering and so no analysis of
the magnetic contribution was then attempted.\par

\begin{figure}
\caption{The neutron scattering intensity observed using MARI as a
function of energy transfer and wave-vector transfer along the
$[001]$ direction. The data shown is an average of measurements at
$6.2$ and $12\;\mathrm{K}$ and is given on a logarithmic scale.
The intense band of scattering arises from the well-defined
Haldane modes and the scattering continuum is at higher energies
up to $\sim 12\;\mathrm{meV}$ for $0.6 < l < 1.4$. The solid lines
are the boundary of the continuum that would be observed for an
$S$=$1/2$ linear chain if the maximum of the lower boundary was at
the same energy as the Haldane excitation for $l=0.5$.}
  \label{MARI_plot}
\end{figure}

Constant-$l$ scans were constructed from the measured
$S(\vec{Q},E)$ by mapping on ($l,E$)-space (Fig.~\ref{MARI_plot}).
This data was then averaged in stripes with a width $\Delta l=0.05
- 0.10$ and corrected for neutron absorption. Fig.~\ref{MARI_plot}
shows directly the continuum scattering. It is stronger at AF
momenta around $l=1$ than in the region near $l=0.2$, which is
known to carry vanishing magnetic weight \cite{Affleck_Weston},
and the zone boundary region $l=0.5$ A detailed description of the
analysis of the experiment will be given elsewhere
\cite{Kenzelmann_CsNiCl3_big}. Here, we focus only on the
scattering close to the AF point as measured in the central
detector bank. As shown in Fig.~\ref{MARI_C5}a) the response at
$Q_{\rm c}=\pi$ consists of the sharp mode observed previously but
also has a component that extends to high frequencies. This
continuous spectrum has not been observed before. It is well above
background and moreover is seen to be much larger than the
scattering near $Q_{\rm c}=0$ where the magnetic scattering is
expected \cite{Affleck_Weston} to be vanishingly small. We will
show below that the continuum cannot arise from resolution
broadening of the well-defined mode.\par

The magnetic excitations in the MARI experiment were best fitted
by an antisymmetrized Lorentzian peak weighted with the detailed
balance factor and convoluted with the line-shape of the
quasi-elastic incoherent scattering, which is a good estimate of
the resolution line-shape particularly at low energy transfers. An
intrinsic Lorentzian peak is supported by a recent theoretical
prediction \cite{Damle_Sachdev} and a recent experiment
\cite{Kenzelmann_CsNiCl3_gap}. It gives an excellent fit to the
low energy resonant Haldane peak. Nevertheless we see in
Fig.~\ref{MARI_C5}a) that the scattering intensity at energies
above the gap energy is significantly higher than expected from
the fits to the sharp peak. The continuum scattering is clearly
observed for $0.6 < l < 1.4$ (Fig.~\ref{MARI_plot}). At
$6.2\;\mathrm{K}$, the integrated intensity of the continuum is
10(4)\% of the total magnetic scattering at the AF point. The
continuum was also observed in an experiment in a different
configuration at $12\;\mathrm{K}$, where the c-axis was parallel
to the incoming beam.\par

Calculations were made using an ISIS spectrometer simulation
programme which takes into account the spectrometer parameters
such as the detailed pulse line-shape, chopper characteristics
etc. and predicts the scattering line-shape for a particular
sample orientation and model of the scattering cross-section. We
find that the resolution function reproduced the measured
line-shape of the quasi-elastic incoherent scattering and of the
well-defined excitations, but that it cannot give rise to a
high-energy tail in the region of the continuum.\par

\begin{figure}
  \caption{a) Neutron scattering intensity at $6.2\;\mathrm{K}$ as
a function of energy transfer at $l=0$ and $l=1$ as measured using
the time-of-flight spectrometer MARI at ISIS and after correction
for neutron absorption. The wave-vector transfer describing the
in-plane wave-vector [hh0] varies between $h=0.09$ and $0.81$. The
solid line is the fit described in the text. b) Neutron scattering
counts at $8.5\;\mathrm{K}$ as a function of energy transfer as
measured using the triple-axis spectrometer DUALSPEC and corrected
for neutron absorption. The energy scan at the zone boundary
$(0.31\;0.31\;1.5)$ shows that the magnetic excitation peak is at
$\sim 6\;\mathrm{meV}$ and that the background at low and high
energy transfer is considerably lower than the neutron scattering
at $(0.31\;0.31\;1)$ between $4$ and $10\;\mathrm{meV}$. The solid
lines are fits described in the text including the resolution
function. The shaded areas are a guide to the eye.}
  \label{MARI_C5}
\end{figure}

The continuum scattering was investigated further using the
triple-axis spectrometer DUALSPEC at Chalk River Laboratories and
constant-$\vec{Q}$ scans were performed at $8.5\;\mathrm{K}$ for
various wave-vectors in reciprocal space. The energy scans close
to the AF point $l=1$ revealed considerable continuum scattering
above the well-defined Haldane excitation. The continuum extends
up to $~12\;\mathrm{meV}$ as shown in Fig.~\ref{MARI_C5}b) for the
energy scan at $(0.31\;0.31\;1)$. This wave-vector was chosen to
be close to the 3D ordering wave-vector such that the Haldane
excitation was at the lowest energy to give the largest possible
energy window for the observation of the continuum. The scattering
intensity at the AF zone boundary, $l=1.5$, below and above the
sharp excitation is considerably lower than the intensity at
$l=1$. The magnetic excitation was then fitted by a dispersion
relation (an antisymmetrized Lorentzian weighted by the Bose
factor) convoluted with the resolution ellipsoid given by
Cooper-Nathans's expression \cite{Cooper-Nathans}. The dispersion
relation has been given by \cite{William_Buyers_1,Shaolong_Ma},
and at $6\;\mathrm{K}$ the zone boundary energy is
$6\;\mathrm{meV}$, the gap at $(0.81\;0.81\;1)$ is
$1.23\;\mathrm{meV}$ \cite{Kenzelmann_CsNiCl3_gap} and the
bandwidth of the gap along $[110]$ direction is given by the gap
at $(0.33\;0.33\;1)$, which is $0.35\;\mathrm{meV}$
\cite{I_A_Zaliznyak}. The Lorentzian half width of the excitation
is $0.35\;\mathrm{meV}$ \cite{Kenzelmann_CsNiCl3_gap}. The fit
gives an excellent account of the right-hand-side of the peak, as
shown in Fig.~\ref{MARI_C5}b), but it cannot explain the slowly
decreasing scattering continuum at energies above
$4\;\mathrm{meV}$.\par

The integrated intensity for the DUALSPEC data at the AF point
$l=1$ was inferred by scaling the peak intensity for $l=1.5$
according to the $l$-dependence of the intensity measured by the
MARI experiment. The integrated intensity of the continuum at
$8.5\;\mathrm{K}$ is found to be about $12(3)\%$ of the overall
scattering at the AF wave-vector. This confirms the results
obtained using the MARI spectrometer.\par

The temperature dependence of the continuum scattering was further
investigated using the RITA triple-axis spectrometer at ${\rm
Ris\o}$ and constant-$\vec{Q}$ scans were performed at the 1D
point \mbox{$(0.81\;0.81\;1)$}. In Fig.~\ref{RITA_scans} we show
that the peak broadens and increases in energy with increasing
temperature as reported elsewhere \cite{Kenzelmann_CsNiCl3_gap}.
The well-defined peak was well fitted by an antisymmetrized
Lorentzian weighted by the Bose factor and convoluted with the
resolution ellipsoid. The flat background was determined from the
scattered intensity below $25\;\mathrm{K}$ at the highest energy
transfers measured ($E>12\;\mathrm{meV}$), where the scattering
was temperature independent and assumed to be non-magnetic. At low
temperatures and energies higher than the well-defined peak
energy, there was considerable continuum scattering
(Fig.~\ref{RITA_scans}). After correction for neutron absorption,
the total integrated intensity at $6\;\mathrm{K}$ and
$12\;\mathrm{K}$ was about 14(3)\% and 13(6)\% of the integrated
scattering at $l=1$, respectively. At higher temperatures, the
Lorentzian peak increases in energy and broadens so that the
continuum could not be clearly identified. Nevertheless there was
more scattering around $8-12\;\mathrm{meV}$ energy transfer at $6$
and $9\;\mathrm{K}$ than at $30\;\mathrm{K}$ (see
Fig.~\ref{RITA_scans}).\par

\begin{figure}
  \caption{Neutron scattering intensity at $l=1$ wave-vector
  transfer as a function of energy transfer measured using the
  triple-axis spectrometer RITA in ${\rm Ris\o}$ for three different
  temperatures. The data is corrected for neutron absorption.
  The solid lines are the fits explained in the text including the
  spectrometer resolution and the shaded areas are a guide to the eye.}
  \label{RITA_scans}
\end{figure}

There are a number of theoretical predictions for the strength of
a continuum at $l=1$. A numerical diagonalization of a spin-1
chain with $N=20$ sites and nearest-neighbor exchange interaction
showed that at zero temperature the continuum above the Haldane
excitation carries $3\%$ of the total weight
\cite{Minoru_Takahashi_3}. In the non-linear sigma model (${\rm
NL\sigma M}$), a continuum arises from 3-particle scattering, and
its integrated intensity is about $2\%$ of the spectral weight
\cite{Horton_Affleck,Essler}. If the coupling of the magnetic
chain is taken into account via a Random Phase Approximation
(RPA), the 3-particle scattering becomes dependent on the
wave-vector transfer perpendicular to the 1D axis, but its
integrated intensity remains between 0.5 and $2.25\%$ of the total
spectral weight \cite{Essler}.

The portion of the intensity in the continuum scattering for ${\rm
CsNiCl_{3}}$ is about $12(2)\%$ for $l=1$. Our measurements for
${\rm CsNiCl_{3}}$ show that the intensity between $l=1$ and
$l=1.5$ reduces by a factor of $5.4$, which is substantially lower
than the theoretical prediction of $8.9$. The absolute intensity
of continuum scattering may therefore be only $7(2)\%$ of the
theoretical predicted intensity, but this is nevertheless much
greater than the theoretical predictions.\par

A considerably stronger 3-particle continuum of about 17\% is
predicted by the Majorana Fermion Theory (MFT), which describes 1D
magnetic systems in terms of a perturbation theory about the
Heisenberg point of a model with biquadratic interactions as
strong as the bilinear exchange \cite{Essler}. It is unclear
whether this model is applicable for ${\rm CsNiCl_{3}}$.
Furthermore, no indication of the pronounced peak at four times
the gap energy, which is predicted for this model, was found in
our measurements. Finally the MFT results are not in agreement
with the numerical calculations of Takahashi
\cite{Minoru_Takahashi_1}.\par

In order to confirm that the continuum scattering arises from
3-particle scattering as predicted by the ${\rm NL\sigma M}$ we
have searched for the onset of the continuum at $3\Delta$ and for
the pronounced maximum of the continuum at $6\Delta$ as predicted
by the ${\rm NL\sigma M}$. With $\Delta=0.94\;\mathrm{meV}$, our
resolution is more than adequate to observe an increasing
continuum intensity between $3\Delta=2.8\;\mathrm{meV}$ and
$6\Delta=5.6\;\mathrm{meV}$, but instead the spectrum shows a
steadily decreasing intensity (Fig.~\ref{MARI_C5},
\ref{RITA_scans}). This may arise because the 3D interactions in
${\rm CsNiCl_{3}}$ cause a change in the excitation spectrum which
is not properly taken into account via RPA and the pronounced
maximum at $6\Delta$ gets much broader as expected from
kinematics. Possibly the effect can also account for the observed
continuum intensity being higher than predicted for a 1D
chain.\par


The observed continuum, Fig.~\ref{MARI_plot}, \ref{MARI_C5},
\ref{RITA_scans}, has intensity extending to approximately
$12\;\mathrm{meV}$, which is twice the maximum 1-particle energy
for ${\rm CsNiCl_{3}}$. Because similar measurements near $l=0,2$
show no evidence of scattering, we conclude that the continuum is
strongest in a broad region where the AF fluctuations are largest.
It is then possibly worth commenting that qualitatively the
continuum scattering has features that are similar to the
continuum found for $S$=$1/2$ systems (Fig.~\ref{MARI_plot})
\cite{D_Tennant}.\par

It is clear that further theoretical work is needed to understand
these results. It is unclear firstly whether a quantitative theory
can be obtained from the 3-particle picture of the ${\rm NL\sigma
M}$ or whether it is more appropriate to use another development
such as the MFT approximation \cite{Tsvelik} to the Heisenberg
model with biquadratic exchange and secondly whether the form of
the scattering depends on the inter-chain coupling or is
characteristic of coupled chains.\par

In summary, we have measured the continuum scattering of the
spin-1 chain compound ${\rm CsNiCl_{3}}$ in a broad region that
favours antiferromagnetic fluctuations and shown that it is
stronger than expected for an ideal spin-1 chain both from
numerical diagonalization and field theoretical calculations
involving the ${\rm NL\sigma M}$. These results are presented as
observations from experiment, and while we cannot explain them we
hope that they will stimulate further theoretical work.

\hspace{0.5cm}

We would like to thank Z. Tun for his assistance and Ian Affleck
and A.~M. Tsvelik for helpful discussions. Financial support for
the experiments was provided by the EPSRC, by the EU through its
Large Installations Program and by the British Council-National
Research Council Canada Program. ORNL is managed for the U.S.
D.O.E. by UT-Battelle, LLC, under contract no. DE-AC05-00OR22725.
One of the authors (M.~K.) is supported by a TMR-fellowship from
the Swiss National Science Foundation under contract no.
83EU-053223.

\end{document}